\documentclass[conference]{IEEEtran}
\IEEEoverridecommandlockouts
\usepackage{cite}
\usepackage{amsmath,amssymb,amsfonts}
\usepackage{verbatim}
\usepackage{algorithmic}
\usepackage{graphicx}
\usepackage{textcomp}
\usepackage{listings} 
\usepackage{xcolor}
\usepackage{subfig}

\makeatletter
\newcommand{\linebreakand}{%
  \end{@IEEEauthorhalign}
  \hfill\mbox{}\par
  \mbox{}\hfill\begin{@IEEEauthorhalign}
}
\makeatother

\def\BibTeX{{\rm B\kern-.05em{\sc i\kern-.025em b}\kern-.08em
    T\kern-.1667em\lower.7ex\hbox{E}\kern-.125emX}}
\begin{document}

\title{Integrating Data Structures and
Algorithms in K-12 Education using Block-based Programming\\
}

\author{\IEEEauthorblockN{Ashwin Jagadeesha}
\IEEEauthorblockA{
University of Massachusetts Lowell\\
Lowell, MA, USA \\
ashwin\_jagadeesha@student.uml.edu}
\and
\IEEEauthorblockN{Pranathi Rayavaram}
\IEEEauthorblockA{
University of Massachusetts Lowell\\
Lowell, MA, USA \\
nagapranathi\_rayavaram@student.uml.edu}
\and
\IEEEauthorblockN{Justin Marwad}
\IEEEauthorblockA{
University of Massachusetts Lowell\\
Lowell, MA, USA \\
justin\_marwad@student.uml.edu}
\linebreakand
\IEEEauthorblockN{Sashank Narain}
\IEEEauthorblockA{
University of Massachusetts Lowell\\
Lowell, MA, USA \\
sashank\_narain@uml.edu}
\and
\IEEEauthorblockN{Claire Seungeun Lee}
\IEEEauthorblockA{
University of Massachusetts Lowell\\
Lowell, MA, USA \\
claire\_lee@uml.edu}
}
\maketitle

\begin{abstract}

This paper describes the design and evaluation of DSAScratch, an extension to Scratch, a widely used block-based programming language. The DSAScratch framework implements advanced data structures such as arrays, sets, dictionaries, and searching and sorting algorithms. By presenting these concepts in an intuitive block-based interface, these blocks abstract away technical details and simplify data structures and algorithms concepts for K-12 students to grasp and apply to programming problems more readily. A preliminary evaluation of the tools' usability and learning outcomes is presented in this paper. Given the information we have gathered about DSAScratch, we show that the extension is beneficial for students to develop a deeper understanding of programming and an intuitive understanding of these concepts in high school. We present the methodology and preliminary results of a user study conducted with ten high school students. During the user study, 70\% of the participants understood the key ideas behind DSAScratch implemented data structures and algorithms through a mixture of lectures and hands-on activities. We show that DSAScratch was also an important part of the workshop for 90\% of the students who participated, as it enhanced their understanding of algorithms and data structures. Furthermore, they indicated that they would recommend DSAScratch to their peers.

\end{abstract}


\section{Introduction}

In recent years, fast-paced innovations in computational technologies have greatly advanced STEM fields. A skilled technical workforce is also in greater demand due to this. There is no doubt that this trend will continue. Many K-12 initiatives and curricula target competencies considered essential for the 21st century, with programming being an essential tool for building these competencies\cite{csimportance1}.
To prepare students for the transition into undergraduate studies
\cite{csimportance2,difficultyteaching2,difficultyteaching3}, it is important that students are taught advanced data structures such as arrays, dictionaries, sets, and their use-cases as part of their computer science education at K-12 level\cite{datastructureimportance1,difficultyteaching2}. Furthermore, the big ideas in computing research recommend that students in grades nine through twelve use data structures, such as lists and dictionaries\cite{fredmartin}. Nevertheless, existing research suggests that covering the essence of such complex topics is more difficult than it appears\cite{difficultyteaching,difficultyteaching2}. Textual programming languages indeed provide these capabilities that are applied in school curricula. However, they cannot accommodate different cognitive abilities\cite{scratchhighschool2,cognition}. This makes understanding the essence of these data structures difficult and short-lived\cite{visuallanguagedatastructures}. To address this, we have developed DSAScratch (Data Structures and Algorithms for Scratch) as an extension to the Scratch programming language. As a teaching tool for introductory programming, DSAScratch helps introduce three data structures: arrays, dictionaries, sets, and a searching and sorting algorithm, all of which have a wide range of applications in programming and computer science education. 
For the development of DSAScratch, we chose Scratch because it is one of the most widely used and popular visual programming tools in the world today \cite{scratchpopular}, popular not only at middle school but also at high school level\cite{AIhigh}. Scratch, Snap, and Blockly, which facilitate learning basic programming constructs, are among the popular tools contributing to this wide adoption in K-12 curricula \cite{scratchwideadoption, scratchimportant}. 

\begin{figure}[!htb]
  \centering
    \includegraphics[width=.46\textwidth]{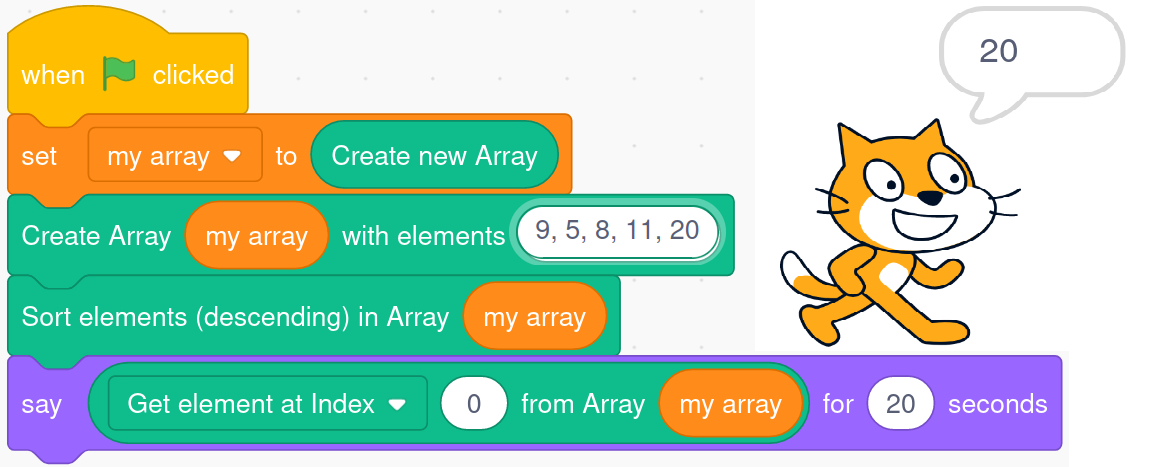}
    \caption{Finding the maximum element in the Array}
    \label{fig:arrayssort}
\end{figure}

Scratch does not currently implement data structures or important algorithms out of the box, which limits its applicability in advanced K-12 high school CS curricula. Through DSAScratch, educators can creatively introduce advanced data structures and algorithm concepts and encourage active learning \cite{activelearning,activelearning1} using a multimedia, visual, and intuitive interface. DSAScratch would offer students an interactive and enjoyable way to learn complex programming paradigms within the Scratch interface \cite{needfortoolsblktotext}.

Several extensions have been developed for Scratch that enables children to learn several advanced concepts, such as Artificial Intelligence, Robotics, Machine Learning \cite{AIhigh,advancedconcepts2}, and Cybersecurity~\cite{cryptoscratch}. These tools demonstrate the usage of Scratch's media-rich environment and its ability to simplify complex concepts to their atomic components. Additionally, visual interfaces can encourage students to develop games and program robots using Scratch\cite{scratchhighschool1}. 


Furthermore, research suggests that students working with block-based interfaces demonstrated greater interest and learning gains in future programming efforts\cite{bridgeprogram,learningtransferblocktotext}. DSAScratch extends the possibilities offered by Scratch as an introductory programming tool to enable students to experiment with advanced algorithms before moving into Python or other more advanced programming languages\cite{scratchbeforepython}. Studies demonstrate that block-based approaches provide a greater advantage than text-based approaches in this context when introducing difficult concepts to novices\cite{scratchhighschool1,novice1}.


\vspace{.5em}
Hence, this research is focused on two interconnected research questions:

\begin{itemize}
\item \textbf{RQ1}: How can we simplify advanced concepts using a visual block-based interface of Scratch to teach complex topics such as data structures and algorithms to high school students?
\item \textbf{RQ2}: Does the simple interface encourage learners to incorporate abstraction, data structures, and algorithms into their projects and to establish connections between basic programming constructs to complex ideas?
\end{itemize}


To answer \textbf{RQ1} we designed the DSAScratch framework and simplified advanced data structures and algorithms into procedurally abstracted blocks within the Scratch environment. 
As part of the Scratch framework, DSAScratch seamlessly integrates data structures and algorithms with existing blocks and concepts within the Scratch environment. The example in Fig. \ref{fig:arrayssort} illustrates how a student can create an array variable and sort the array in descending order to find the maximum element in the array. The index for this element is either at the beginning or the end of the array, depending on whether you choose to sort in ascending or descending order. By combining an array data structure with a sorting algorithm, the example illustrates how to offer an intuitive solution to the problem of finding the maximum element in the array. This provides learners an insight into how indices can be used to arrange items inside an array. In this way, abstract blocks are atomic steps in the same sense as other Scratch blocks.

Further, to answer \textbf{RQ2}, we evaluated DSAScratch's ability to simplify complex ideas, as well as its ability to teach intuitively. Therefore, we introduced it to a group of ten high school students with little experience with programming languages such as Python, C, C++, and Scratch and were all novices at programming. Consequently, we introduced the concepts of data structures and algorithms as part of a  three-day workshop through the DSAScratch framework on the Scratch interface. As part of our evaluation, we administered two quizzes. In Quiz-1, students are assessed on their ability to differentiate basic programming constructs in Scratch. In Quiz-2, students were assessed on their awareness of advanced data structures and algorithms covered during DSAScratch demonstrations. In addition, three laboratory challenges were designed to test how these ten students could apply  concepts learned in the workshop to solve lab challenges by combining DSAScratch concepts with basic constructs in Scratch. As part of the lab challenges, students applied various algorithms and data structures taught during the 3-day workshop.
According to the post-workshop survey, 90\% of the students felt positive about learning data structures and algorithms. Among the participants, 80\% rated DSAScratch as 5 and 20\% as 4 on a Likert scale (1-5), indicating a general liking towards DSAScratch. Over 90\% of the participants said they would recommend DSAScratch to their friends. 
A detailed evaluation of the framework is presented in this paper.

By focusing on K-12 high school education, we test the framework's ability to simplify instruction of such concepts as the evaluation of DSAScratch's effectiveness and accessibility to K-12 high school students. With DSAScratch, students learned how to apply data structures on the Scratch platform and learned about their applicability to solving real-world use cases. In addition, the tool demonstrates its ability to scaffold learning \cite{scaffoldedlearning} as part of the Scratch interface.

\section{Related Work}

Research shows that several tools have been developed to make advanced programming ideas more accessible to young learners\cite{AIhigh,advancedconcepts2,deepscratch,cryptoscratch,blocksforsensors,advancedconceptsscratch1}. Block-based learning methodologies are widely used in K-12 computer science educational context as these tools address difficulties associated with learning syntax in textual programming languages\cite{novice1}. Many K-12 CS tools make use of simulation and gamified learning computer science concepts\cite{lowcost,scratchhighschool1}. Visualizations and simulations make complex concepts easier to understand and cater to the different cognitive abilities of young students. Visual programming environments such as Scratch, Blockly, and Snap provide block-based interfaces based on this philosophy. This research  \cite{difficultylearningDS} raises awareness of the difficulties students face when learning data structures like ArrayList, LinkedList, and Trees in textual programming languages but provides no tools in particular. In addition, the authors recommend that the CS education community explores this topic further, noting that there is still a limited amount of literature on the topic.
Topics such as data structures require a high level of abstraction. Research suggests that scaffolded learning tools are necessary \cite{dstblocks,block4ds}. One of the main advantages of scaffolded instruction is that it fosters an environment of supportive learning. Students contribute their knowledge by asking questions, providing feedback, and helping their peers learn new material \cite{block4ds}. Environments such as Scratch foster Scaffolded learning opportunities for students. DStBlocks \cite{dstblocks} is an example of a scaffolded visualization tool that extends Blockly to simulate advanced data structures. There is block4ds, a block-based simulation tool developed to teach undergraduate computer science students advanced data structures such as Binary Search Trees (BST) \cite{block4ds}. These tools are designed to help students visualize hard concepts and help view data structure manipulations by generating relevant visual models representing the data structure, such as generating a BST after a node has been removed from the tree. These tools are aimed at undergraduate students.

DSAScratch, on the other hand, is designed to help K-12 students explore data structures and \textit{apply them} to actual use cases in a block-based programming context. DSAScratch has a strong integration as an extension into the Scratch framework, which allows students to create programs that incorporate such concepts into learning. Furthermore, it complements the current Scratch interface in its ability to view data structures as variables in Scratch, this allows users to view manipulations on data and inspect resulting changes to a data structure within the Scratch interface. It aims to help students formulate good solutions to problems and develop high-level abstraction capabilities while writing programs. Thus, when students switch to textual languages, they possess some familiarity with such advanced concepts.

\section{Methodology}
\label{framework}

In this section, we focus on providing an overview of the DSAScratch framework. 
Section \ref{choiceofconcepts} covers the motivation behind the concepts chosen for the proof of concept on Scratch for the initial set of data structures and algorithms. 
Section \ref{design} covers the three design principles behind the creation of the framework. Furthermore, Section \ref{bati} provides a detailed description of how DSAScratch blocks function. In addition, it discusses the applicability of advanced concepts within the Scratch programming environment.

\subsection{\textbf{Choice of concepts}}
\label{choiceofconcepts}

\begin{figure}[t]
  \centering
  \begin{minipage}[b]{0.5\textwidth}
    \includegraphics[width=\textwidth]{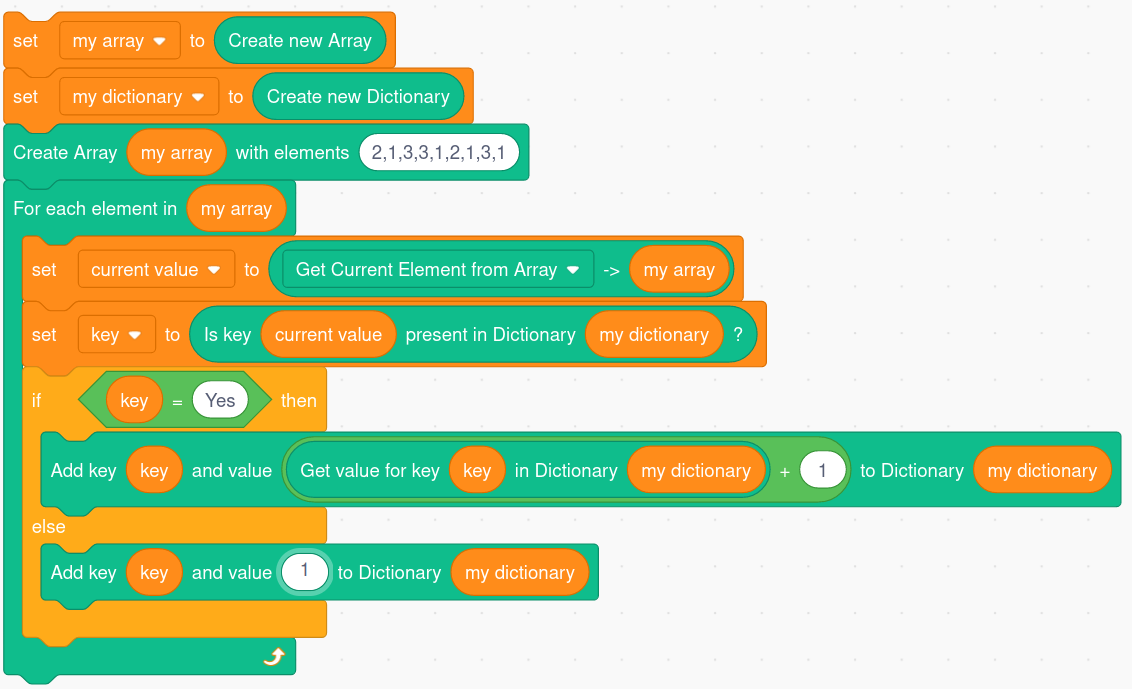}
    \caption{Frequency of elements in the array}
    \label{fig:dictionaries}
  \end{minipage}
\end{figure}


In programming languages, complex data can be represented as various data structures that can be accessed and manipulated efficiently. DSAScratch currently supports three data structures: Array, Set, and Dictionary, and one searching (internally implements Binary-search) and sorting algorithm (internally implements Quick-sort).
During the selection process, data structures and algorithms were assessed for their applicability to programming contexts. As our preliminary selection, we chose the following concepts.

\vspace{0.2em}
~\textbf{Arrays:} Arrays are data structures that store a collection of items. Aside from primitives, arrays can also hold references in textual programming languages. In terms of storing and retrieving a sequence of items, arrays are the most efficient data structure. Arrays are ubiquitously used in programming. Scratch currently offers a List data structure for creating a collection of items. DSAScratch arrays are similar in functionality to these blocks. The DSAScratch array, however, has been designed to work with other Algorithms and data structure blocks within the DSAScratch framework. With DSAScratch arrays, it is possible to use the sort by ascending or descending blocks to organize data within arrays, and the search block can be used to perform searches within the array data structure. Scratch lists are not well suited for achieving this kind of goal. 


\vspace{0.2em}
~\textbf{Set:} A Set data structure can contain any number of unique values in any order you choose. In contrast to arrays, sets contain only non-repeated, unique elements. It is essential to understand Set Theory and Sets in mathematics, and Sets have a lot of relevance to computer science through its applications. An important property of a set is that it can be operated using operations such as \texttt{UNION}, \texttt{INTERSECTION}, and \texttt{DIFFERENCE}. Sets provide O(1) inserts and O(1) lookups making them highly efficient and utilized in programming and computer science education. An algorithm's performance can be greatly affected by the data structures it uses. The set data structure is ideal if you create an item set composed solely of unique items.   

\vspace{0.2em}
~\textbf{Dictionaries:} A Dictionary is an associative data storage container that maps a key to a value which can be defined as  the mapping of two different components that share a relationship. Hence, it stores data as a collection of key-value pairs. It is possible for a dictionary to act as a set when both the key and value represent the same item. To illustrate their relevance in real-world programming situations, it is important to distinguish between a set and a dictionary.
A Dictionary is particularly useful when looking up and tallying data elements quickly to find the number of students in UG vs. PG in a university. For example, consider a dictionary where the student's name is the key and the degree level is the value associated with this key \texttt{Student information} = \{\textit{Aaron: UG, Emily: PG}\}, a dictionary-based key-value lookup would be ideal if you want to count the number of undergraduates vs. postgraduates quickly, the solution can be reduced to O(n) using a dictionary. Similarly, there is moderate difficulty in finding a solution to a problem, like counting the frequency of each element in an array with repeated items. It is, however, possible to greatly simplify the solution to this problem by using a dictionary. We can design an intuitive solution by using a dictionary to store an element as a key and its frequency as a value associated with the key. When a repeated element is encountered, dictionaries can check if the key already exists. Whenever the key matches, increment the value stored for the key referring to Fig. \ref{fig:dictionaries}.


In this sense, DSAScratch is capable of solving challenging programming puzzles that go beyond simple examples. Programmers will benefit from the ability to derive key-value relationships from data. In addition, this leads to programs that manage resources efficiently.

\begin{figure}[t]
  \centering
    \includegraphics[width=0.5\textwidth]{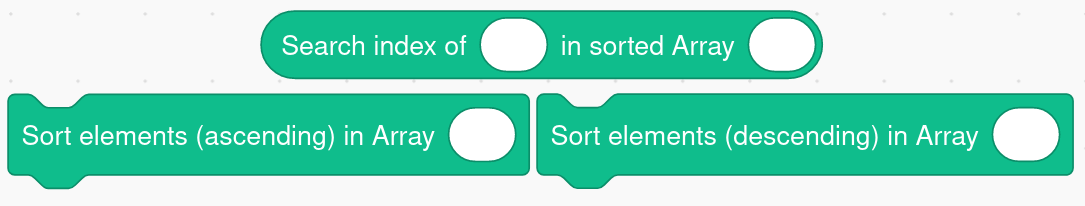}
    \caption{Algorithm blocks}
    \label{fig:algoblocks}
\end{figure}

\vspace{0.2em}
~\textbf{Search and Sort Algorithms}: Sorting algorithms describe how data should be arranged and stored. Ordering data makes searching more efficient. Besides retrieving information and searching virtual spaces, it can be used to identify useful patterns in data. In the most common cases, data is sorted numerically or lexicographically. A numerical order describes the arrangement of the data in ascending or descending order. A lexicographical order defines the overall order over sequences of text that are arranged alphabetically. All data structures in DSAScratch are designed to accept text and numerical input. Furthermore, the sorting blocks can sort numerically and lexically, and the Search block can search numerical and textual data within an array and return indices of matching items. Referring to Fig. \ref{fig:algoblocks} for searching and sorting blocks.

\begin{figure}[t]
  \centering
  \begin{minipage}[b]{0.4\textwidth}
    \includegraphics[width=\textwidth]{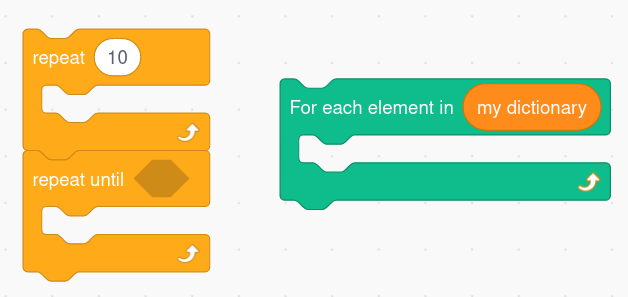}
  \end{minipage}
  \caption{Scratch Loop blocks (left), DSAScratch block (right)}
  \label{fig:loops}
\end{figure}

\vspace{0.2em}
\textbf{Looping within DSAScratch data structures:} We created a new looping block that is compatible only with DSAScratch's data structures to avoid interfering with Scratch's existing loop functionality. The \texttt{for each element in} block accepts a data structure as input and helps iterate within it as shown in Fig. \ref{fig:loops}. 

\subsection{\textbf{Design principles behind DSAScratch}}
\label{design}

\begin{figure}[t]
  \centering
    \includegraphics[width=0.43\textwidth]{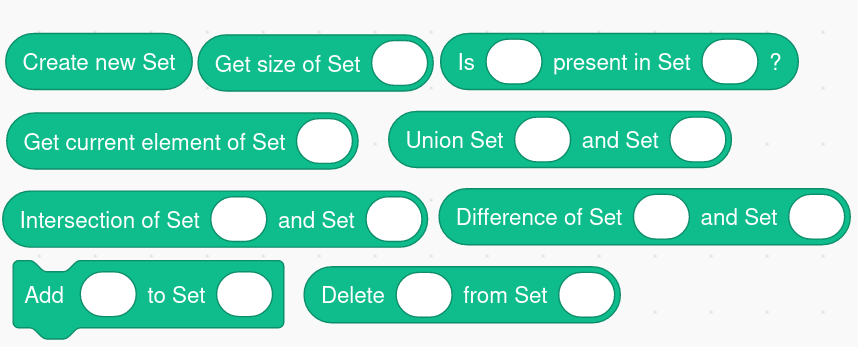}
    \caption{Blocks for Set}
    \label{fig:setsfinal}
\end{figure}

DSAScratch conforms to the core design principles behind Scratch, which is to make it more accessible, and more tinkerable \cite{Scratchmain}. Hence, the proposed design can be categorized into three design ideas. 

\vspace{0.2em}
\subsubsection{\textbf{Procedural abstraction of DSAScratch blocks}}
\label{paab}
Our goal is to abstract blocks with an appropriate level of complexity without losing sight of the core functions of a data structure or algorithm. Abstraction of procedures can help remove code smells\cite{atomiccomponentsprocedure}. The concept of procedural abstraction involves decoupling the logical properties of an operation from the details of its implementation. The most common method of accomplishing this is by moving fragments of code into a procedure\cite{atomiccomponentsprocedure}. Consequently, the blocks balance technical detail and abstraction without losing sight of their function. For example, let us consider the set data structure blocks shown in Fig. \ref{fig:setsfinal}. Sets store unique elements without any particular order. In DSAScratch, an operation such as providing the common items from two sets can be procedurally abstracted to a single block as an intersection operation. Fig. \ref{fig:setsfinal} row 3 block 1 implements this block. Thus, it is possible to develop solutions to problems that call for this idea. This enables users to connect concepts on a high level and deal with more challenging problems, thus incorporating high-level abstraction abilities in their Scratch programs.

\vspace{0.2em}
\subsubsection{\textbf{Creation of extensible blocks}}
\label{coeb}

\begin{figure}[t]
  \centering
  \begin{minipage}[b]{0.5\textwidth}
    \includegraphics[width=\textwidth]{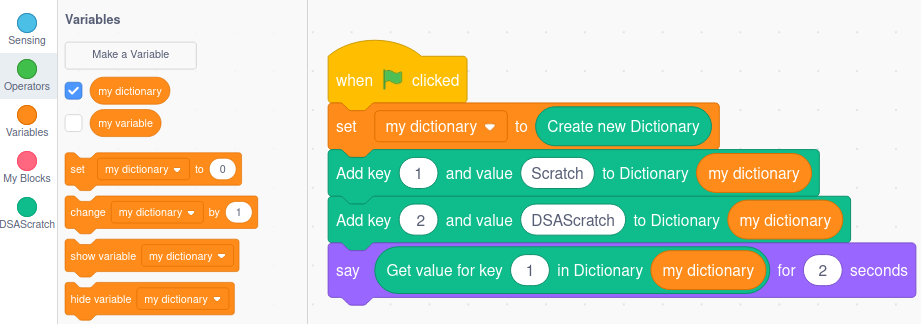}
    \caption{Creation of DSAScratch dictionary variable in Scratch}
    \label{fig:variable}
  \end{minipage}
\end{figure}

DSAScratch's extensibility with default Scratch blocks was a key challenge during its design. We believe it is important for new blocks implemented as a part of DSAScratch to integrate seamlessly with other Scratch block categories\cite{blkcategories}. DSAScratch carefully considers this during its design. Users can create variables in Scratch by using the "Make a variable" block under the variables category of Scratch blocks. With the capability of extending Scratch's "Make a variable" feature to support new data structure blocks referring to Fig. \ref{fig:variable}. For example "create new Dictionary" block can be used to create a dictionary data structure as illustrated in Fig. \ref{fig:variable} and further \texttt{Add key and value} block can be used to add a key-value pair into this newly created DSAScratch dictionary variable. This function aids in bridging existing Scratch blocks with DSAScratch data structures and algorithms blocks. Therefore, Scratch's "Make a variable" feature can now accommodate data structures as variables.

\vspace{0.2em}
\subsubsection{\textbf{Conversational style naming of blocks}}
\label{csnob}
A variety of new blocks have been created, whose names are created in a conversational style. Generally speaking, a conversational block resembles a conversation that occurs between friends. For example, if we want to check if a Set data structure is empty or not, we will use the block ‘‘is set empty?’’. As a result, the answer would either be yes or no. By executing and inspecting the results of conversational blocks, students can gain familiarity and knowledge of their use\cite{toee}.

\section{Blocks and their Functions}
\label{bati}

\begin{figure}[t]
  \centering
    \includegraphics[width=0.5\textwidth]{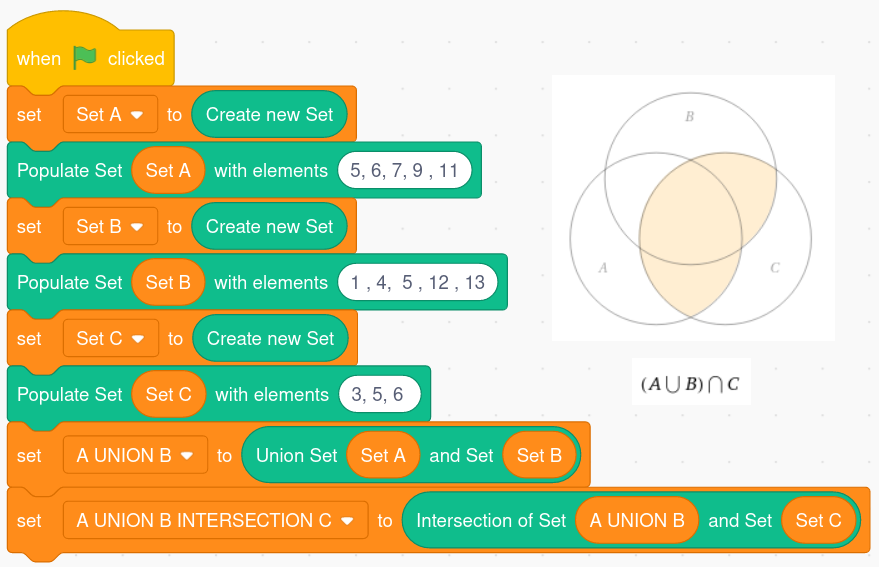}
    \caption{Examples of Set operations that can be performed}
    \label{fig:setsoperations}
\end{figure}

Our framework's main objective is to provide Scratch users with full-featured implementations of common data structures and sorting and searching algorithms as blocks that can function atomically on DSAScratch data structures created as variables within Scratch environments. To achieve this, we carefully considered the existing block functions in Scratch. Existing block functions in Scratch can be categorized into \texttt{COMMAND} block that does not return any result, \texttt{C} block that controls the running of the program, \texttt{REPORTER} block that reports a value after execution \cite{blockstypes}. A \texttt{COMMAND} or otherwise called \texttt{STACK} block is designed so that it can be connected (stacked) with other blocks.

In DSAScratch, all data structures implement \texttt{COMMAND} and \texttt{REPORTER} block types; an example of these block types can be referred to in Fig. \ref{fig:setsfinal}. A \texttt{COMMAND} block does not return any value. Instead, the core function of this block is to execute a command. In DSAScratch, blocks of this type are used to add elements into a data structure or order data within the array using sorting blocks (which are designed as \texttt{COMMAND} blocks) referring to Fig. \ref{fig:algoblocks} row 2. 

Next, Blocks that control the execution of a program have notches at the top and bumps at the bottom. In addition to placing blocks above and below other blocks, it is also possible to place blocks inside them. Blocks like these are called \texttt{C} blocks, \texttt{forever}, and \texttt{repeat} blocks in Scratch are examples of this block type referring to Fig. \ref{fig:loops} left part. The \texttt{for each element in} block in DSAScratch extends this concept to loop over a collection of items within a DSAScratch implemented data structure referring to Fig. \ref{fig:loops} right.

DSAScratch carefully incorporates these ideas when designing block functions to extend the concept of Scratch blocks to data structures and algorithms. Fig. \ref{fig:variable} illustrates the variable creation process in Scratch. In DSAScratch, variable creation blocks such as \texttt{create new array}, \texttt{create new set}, and \texttt{create new dictionary} are implemented as \texttt{REPORTER} blocks. The blocks return a string containing a random identifier linked to an object through which variables can be referred to. DSAScratch integrates with Scratch's \texttt{Set variable name} feature of Scratch. This way, DSAScratch data structures can be created as variables in Scratch. To create a data structure, first, a variable block with the preferred data structure must be created. This allows blocks implemented as operations to be executed on this created variable, such as \texttt{"Add Key and Value to Dictionary"} as described in Fig. \ref{fig:variable}.

In cases where a block's operation needs to report its results, \texttt{REPORTER} blocks are used. In the case of Sets, DSAScratch can provide new Set objects as a result of an operation such as \texttt{"Set A UNION Set B"}. This new Set object can create new Set variables on the fly as Scratch variables. Therefore, these newly created Set variables can be used to perform more set operations. Any number of sets or variables can be compared by operations such as \texttt{UNION, INTERSECTION, and DIFFERENCE} Fig. \ref{fig:setsoperations} illustrates this idea with a Venn diagram for three sets. It is possible to find \texttt{"Set A UNION Set B INTERSECTION Set C"}. As a result, DSAScratch can support the rules of Set theory. This is possible as the \texttt{UNION, INTERSECTION, and DIFFERENCE} blocks are implemented as \texttt{REPORTER} blocks. This capability of DSAScratch Sets allows learners to apply basic mathematical theory to computer science problems through the usage of Set data structure blocks. \texttt{COMMAND} blocks do not return any results, such as adding an item to Set Fig. \ref{fig:setsfinal} row 4 block 1. A user can visualize the contents of a data structure on the Scratch palette once an item has been added to a data structure. The variables in DSAScratch are, therefore, consistent with those in Scratch. We note that the same idea is considered for both dictionaries and arrays. In DSAScratch, data structure blocks are grouped to reduce ambiguity. Whenever an incorrect block not related to a data structure is used to operate on a data structure, for example, when a user tries to use a dictionary \texttt{REPORTER} block or \texttt{COMMAND} blocks with an array block, DSAScratch displays an error message \texttt{"This block can only be used with Arrays"}. As a result, it helps the learner identify what block to choose depending on the data structure they are trying to operate on.

\section{Design of the Framework}
\label{sec:desblocks}
Scratch is transpiled into ES6, which is compatible with most modern browsers\cite{es6}.
To ensure compatibility with Scratch, the DSAScratch framework has been developed following the guidelines for creating new extensions in Scratch. To further ensure that DSAScratch is compatible with modern browsers, our framework design focuses on implementing wrappers for data structures and algorithms already built-in within JavaScript.


We also emphasize that our goal with implementing the blocks is to make them intuitive and easy to use. 
In terms of implementation within Scratch, all the data structures and algorithms have been implemented as JavaScript functions using Scratch guidelines. We use Sets to demonstrate the design and implementation, however, we note that the discussion applies to all the blocks within the framework.
Let's consider a block that checks whether an element is present in a set (see Fig. \ref{fig:setsfinal}, Row 1 block 3) whose code representation is shown in Listing 1.
Here, the block is represented using the text \texttt{"Is [ELEMENT] present in Set [OBJ\_ID]?"} and implemented as a \texttt{REPORTER} block type. This \texttt{REPORTER} block accepts two inputs \texttt{[ELEMENT]} (the element to be checked) and \texttt{[OBJ\_ID]} (the set object) and outputs a single value \texttt{YES/NO} depending on whether the element exists in the set or not. We note that the object id \texttt{[OBJ\_ID]} is a unique random value that is assigned to an object when its constructor is invoked, e.g., when the \texttt{Create new Set} block is executed in our case. This is designed to overcome a restriction in Scratch where objects cannot be directly returned from \texttt{REPORTER} blocks, and the object id is used as an alias. 
The \texttt{opcode} specifies the JavaScript function (\texttt{checkContainInSet}) to be executed on the invocation of the block. Internally, the \texttt{[ELEMENT]} and \texttt{[OBJ\_ID]} values are provided directly as arguments to the function, i.e., \texttt{checkContainInSet([ELEMENT], [OBJ\_ID])}.

\lstset{
   extendedchars=true,
   basicstyle=\scriptsize\ttfamily,
   showstringspaces=false,
   showspaces=false,
   numbers=left,
   numberstyle=\scriptsize,
   numbersep=8pt,
   tabsize=2,
   breaklines=true,
   showtabs=false,
   captionpos=b
}

\vspace{0.5em}
 \begin{lstlisting}[frame=single,basicstyle=\small,caption={Internal Representation of a Set Block},captionpos=b]
opcode: 'checkContainInSet',
blockType: BlockType.REPORTER,
text: 'Is [ELEMENT] present in Set [OBJ_ID]?'
 \end{lstlisting}
 

The \texttt{checkContainInSet} function then invokes the internal Set operations to check the presence of the element \texttt{[ELEMENT]} within the set. Another example from sets that use a different block type \texttt{COMMAND} is for adding an element to the set (see Fig. \ref{fig:setsfinal}, Row 4 block 1). In this specific case, a Scratch \texttt{COMMAND} block is more appropriate as the add operation need not return a value. The DSAScratch blocks also implement error checking to ensure that the framework does not crash in instances where the learner makes an error. For example, performing the check element operation on an empty set will return a message ``Invalid Set'' to alert the learner that the specific object has not been instantiated.

\section{Preliminary Evaluation and Results}\label{AA}

A detailed description of the user study design, the recruitment strategy, the evaluation of the framework, and the learning outcomes achieved by ten high school students is presented in this section.

\subsection{\textbf{Design of the User Study}}
We conducted a 3-day virtual workshop with ten high school students. During this workshop, we examined how DSAScratch impacted students’ understanding of data structures and algorithmic thinking and their ability to apply these concepts. Below are the workshop details, the participants, and the study. The user study has been approved by the university's Institutional Review Board (IRB).

\vspace{0.2em}
\subsubsection{\textbf{Recruitment method}} A snowball sampling method was used for recruitment, beginning with the immediate contacts of the research team and working our way outward. Using the above method, 14 High school students registered for the workshop, out of which 10 students attended. Besides the pre-workshop survey, we administered 2 quizzes and 3 challenges, and a post-workshop survey. The students also submitted their answers to 3 challenges.
The students who participated in the workshop were from the 11th and 12th grades. The gender breakdown of the students included 70\% male and 30\% female students.


\vspace{0.2em}
\subsubsection{\textbf{Pre-training survey}} 
\label{presurvey}
Before the workshop, we asked students to provide a pre-training survey to share their demographic information, programming comfort level, experience with Scratch, and their prior interaction with data structures and algorithms. Quantitative information from the Pre-survey results revealed that 3 students (30\%) had no programming experience, and 7 students (70\%) had limited experience. Furthermore, (70\%) of the students were only able to use simple programming constructs like conditionals and loops. Three students (30\%) had used Scratch previously, while seven (70\%) had never used Scratch before the workshop.
Most students (80\%) had no experience with complex data structures or algorithms. In other words, only 2 students (20\%) had any experience with data structures before the workshop, both students mentioned that they knew about arrays but not about dictionaries or sets. The same applies to sorting or searching. 70\% of the students expressed their interest in studying computer science in the future, and the rest (30\%) answered maybe. Pre-survey results indicate that students who participated in the workshop appeared interested in learning more about programming and computer science for future endeavors in this domain.

\vspace{0.2em}
\subsubsection{\textbf{Training / Workshop}} 

The training workshop was conducted over three days for 9.5 hours divided with 3 hours of activity on Day-1 and 3.5 hours on Day-2, and 3 hours on Day-3. During Day-1, students learned how to write loops, use conditionals, and use conditionals within loops in Scratch, as well as basic input-output (I/O) blocks. As we covered the nine categories of Scratch blocks, we paused between concepts to allow students to develop their programs based on the key constructs from each category. As a culmination of the first day of learning, students wrote basic Scratch programs during a 1-day recess between Day-1 and Day-2. We answered questions about Day-1 concepts during recess. We began Day-2 with a 15-minute quiz on Scratch programming constructs via Quiz-1 and returned scores immediately for students to review. To ensure students understand Scratch blocks before moving on to DSAScratch.  Day-2 of the workshop focused on data structures and algorithms through the DSAScratch framework using simple examples and a hands-on activity where students applied data structures and algorithms to simple Scratch programs. There were many questions about concepts taught using DSAScratch during this period.

Additionally, we prepared video clips demonstrating the usage of data structures and algorithms blocks implemented in the DSAScratch framework for students to refer to during a 1-day recess after Day-2. Day-3 began with an hour-long review of Day 2 lectures, followed by Quiz-2 to test their understanding of Data Structures and Algorithms fundamentals taught as a part of the AlgoScratch extension, which was 30 minutes long. We returned Quiz-2 scores immediately, as this could help them review their answers and clarify concepts with the instructors. To help students clear concepts and further solidify their understanding of data-structures concepts, another hands-on programming session to clarify concepts related to Scratch and DSA Scratch, followed by 30 minutes of Q\&A. Post this, as a next step, we gave students three lab challenges that were easy to moderate in difficulty. The challenges were due in 1.5 hours. Some students requested more time because they believed lab challenges were tractable and could be solved within a day. In this sense, the DSAScratch interface seems to have engaged the students. Participants were sent a post-survey questionnaire after completing the lab challenges.

\vspace{0.2em}
\subsubsection{\textbf{Quizzes and Challenges}} 
\label{quizzes}

\begin{figure}[t]
  \centering
    \includegraphics[width=0.5\textwidth]{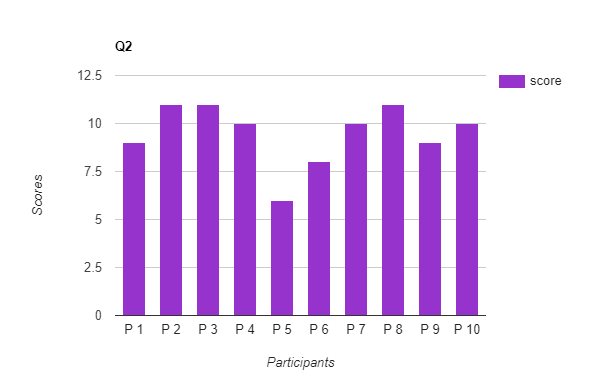}
    \caption{Quiz-2 scores of 10 participants}
    \label{fig:quiz2}
\end{figure}

Students were required to take two multiple-choice quizzes to test their understanding of the topics covered during the lecture. Through Quiz-1, Our objective was to assess students' understanding of basic block concepts in Scratch. 
Therefore, Quiz-1 asked students to distinguish between block categories covering basics such as (I/O), looping, controls, and conditional blocks. Students needed to understand Scratch because all challenges required DSAScratch and Scratch blocks. In addition to answering any questions students had offline after Day 1 of the workshop, we encouraged them to write simple programs that would make the sprites interactive to help students build familiarity with the Scratch platform. They were quizzed on how well they understood Algorithms and Data Structures after learning them with DSAScratch. During Quiz-1, students were evaluated on their ability to differentiate between blocks and concepts within the Scratch interface; however, Quiz-2 was designed to assess their understanding of data structures and algorithms. This was a test to determine if the students understood the key concepts of data structures like arrays, sets, and dictionaries and the fundamentals of each of these data structures.

An example of the question asked in Quiz-1: 
\begin{itemize}
    \item \textit{Which Scratch blocks will you use to perform conditional steps in your program?}
\end{itemize}
\vspace{0.5em}

Similarly, Quiz-2 was designed to test the fundamentals of data structures and algorithms taught using DSAScratch. 

\begin{itemize}
    \item \textit{Consider a Dictionary D = \{1: Potato, 2: Milk, 3: Honey\}. If we add a key-value pair to a Dictionary \{2: Yogurt\}, What would the Dictionary look like after adding the pair?}
\end{itemize}
\vspace{0.5em}

At the end of Day-3, we asked students to complete three coding challenges to final check their ability to apply concepts learned to programs in a hands-on lab session after the workshop.
\begin{figure}[t]
  \centering
    \includegraphics[width=0.5\textwidth]{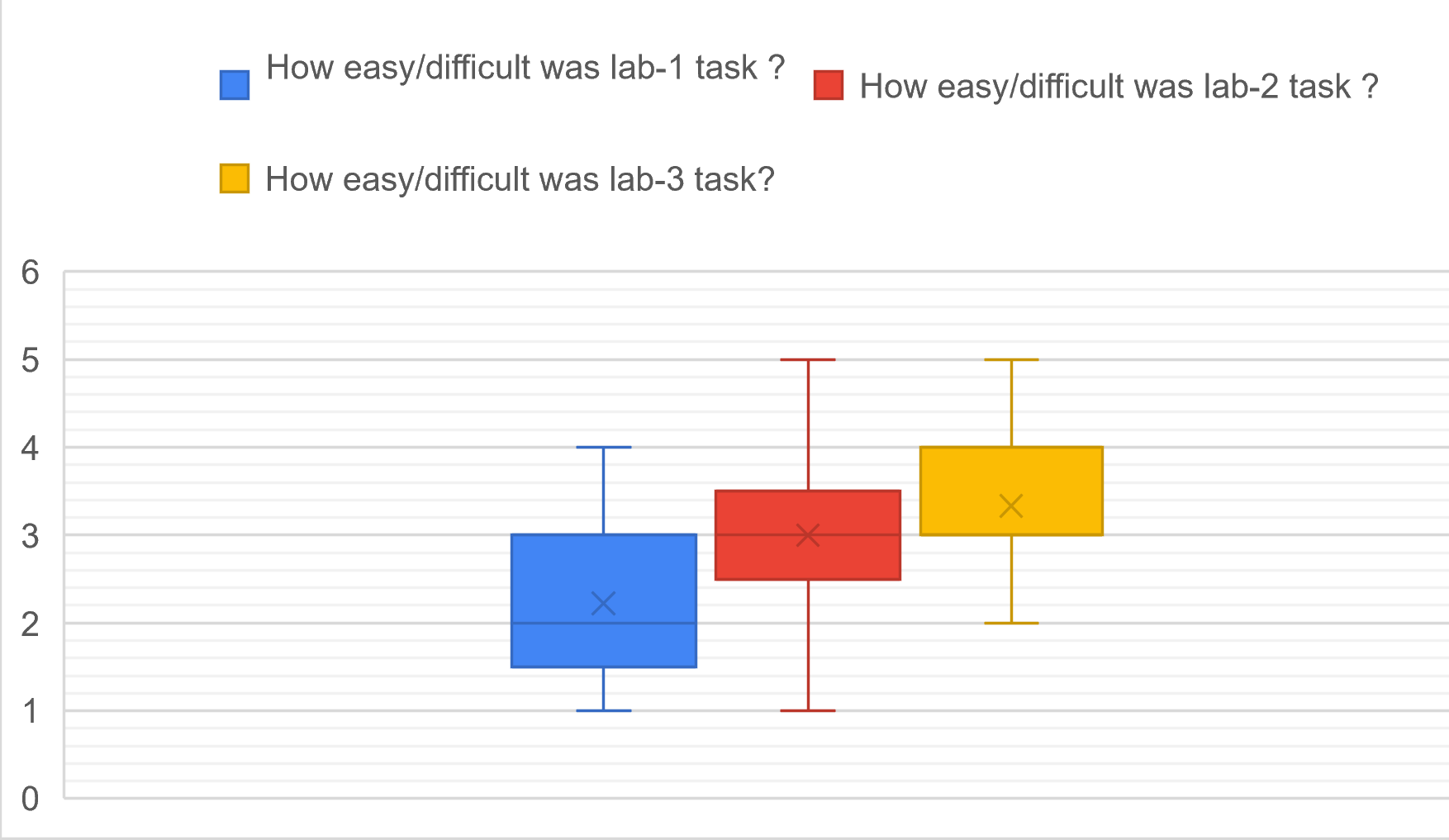}
      \vspace{-1em}
    \caption{Difficulty of Lab challenges as rated by students (1-5)}
    \label{fig:labdiff}
\end{figure}
\begin{itemize}
    \item \textbf{Lab-1}: Can you write a program that accepts the names of fruits and stores them in a DSAScratch Data-structure Array? Can you also display the names of all the fruits stored in this Array? Next, can you sort the names in Ascending order and display each item? Finally, Search if a fruit exists inside that Array. If found, make the sprite say ``yes'' else, say ``not found''.

    \vspace{0.2em}
    \item \textbf{Lab-2}: Can you create  2 different set variables and input items (Example: names, numbers, etc.)? Next, find the UNION of items in both sets and make the sprite display these words for 2 seconds. Finally, find the difference in items in both sets and make the sprite display the difference for 2 seconds.

    \vspace{0.2em}
    \item \textbf{Lab-3}:
    Create a dictionary of words and their meanings. Accept these words and their meanings from the user.
    The program should search a word from the dictionary you created. Accept input search word from the user and get the word's meaning and make the sprite display the word's meaning searched for 20 seconds.
\end{itemize}

In \textbf{Lab-1}, the students were tested on their ability to use array concepts by applying basic Scratch (I/O) constructs. Furthermore, we wanted to test their understanding of ordering and searching. Check if students can handle the case where the item isn't found and be able to report it. With \textbf{Lab-2} To see how well students understand Set data structures, we wanted to see if they could create set variables and perform UNION and intersection on them—building on concepts from \textbf{Lab-1} in \textbf{Lab-2} to work with set variables. \textbf{Lab-3} was designed to check if students could create a mapping between a word and its meaning and use a dictionary to model the problem. Also, whether they can perform lookups for data inside the dictionary.

Each day, we attempted to provide students with enough information about data structures to inspire them to apply them to real-world use cases. Students submitted solutions to three lab challenges we administered after the workshop. Through student solutions, we observed the application of concepts such as ordering items, searching, and efficient storage and covered most of the data structures and algorithms discussed in the workshop. 



\vspace{0.2em}
\subsubsection{\textbf{Post-training Survey}}
As part of the post-training survey, students evaluated DSAScratch, the difficulty of the challenges they solved referring to Fig. \ref{fig:labdiff}. In addition, how effective DSAScratch was at solving these Lab challenges. Lastly, their experience and motivation to use DSAScratch in the future. The post-training survey aimed to understand the tool's efficacy, accessibility, and contribution to students' learning. We wanted to determine if students enjoyed using the tool. Did DSAScratch motivate students to gain conceptual understanding and explore the practical usage of algorithms and data structures?

\vspace{0.2em}
\textbf{\textit{Post Survey Results}}: To evaluate the framework, quantitative and qualitative views are taken. At this stage, it is difficult to draw any statistical inferences from the framework due to the small sample size. Because this was a preliminary evaluation of the DSAScratch framework, our primary goal was to examine the system's ability to motivate high school students as an initial step before performing further evaluations with various demographics. Post-survey questions were designed to infer how students felt about DSAScratch and the experience of learning these concepts in a workshop setting. 
\begin{table}[t]
\centering
\caption{Quantitative Results of Post Survey}
\label{tab:my-table}
\begin{tabular}{|l|l|l|}
\hline
ID & \multicolumn{1}{c|}{\textbf{Question}}    & \multicolumn{1}{c|}{\textbf{Results}} \\ \hline
Q1 &
  \begin{tabular}[c]{@{}l@{}}Did DSAScratch help you solve\\ the questions asked?\end{tabular} &
  \begin{tabular}[c]{@{}l@{}}80\% rated 5/5\\ 20\% rated 4/5\end{tabular} \\ \hline
Q2 & Did the video snippets help you?          & 100\% rated 5/5                      \\ \hline
Q3 & Rate the difficulty of the lab challenges & Refer Fig. \ref{fig:labdiff}                       \\ \hline
Q4 &
  \begin{tabular}[c]{@{}l@{}}Would you recommend DSAScratch as \\ a part of Scratch to your peers ?\end{tabular} &
  \begin{tabular}[c]{@{}l@{}}90\% answered Yes\\ 10\% answered Maybe\end{tabular} \\ \hline
Q5 &
  \begin{tabular}[c]{@{}l@{}}Do you plan to use DSAScratch \\ as part of Scratch for your \\ programming projects ?\end{tabular} &
  \begin{tabular}[c]{@{}l@{}}70\% answered Yes\\ 30\% answered Maybe\end{tabular} \\ \hline
\end{tabular}
\end{table}

To assess DSAScratch quantitatively, we asked the following questions refer to Table. \ref{tab:my-table}. \textbf{Q1:} To derive whether DSAScratch was easy to use and sufficient for completing the lab challenges. \textbf{Q2:} Students were tested to determine if they could learn the usage of DSAScratch blocks during recess. The only supporting material was provided in the form of video clips demonstrating different blocks for reference. \textbf{Q3:} Were the students able to gauge the difficulty of the tasks given that all of the material was covered? \textbf{Q4:} To derive if the tool has been positively received. \textbf{Q5:} To derive whether students would use DSAScratch for future learning. As a next step, students' opinions regarding the DSAScratch framework were collected through qualitative post-survey questions.
\textbf{Q1:} Describe a data structure you liked the most and why? 70\% answered Dictionary a data structure currently unavailable on the Scratch interface and 10\% answered Arrays, 10\% answered Sets and 10\% did not choose to answer. To gain insights into whether students liked solving problems using data structures. \textbf{Q2:} In your opinion, what is the strength and weakness of DSAScratch? 
To obtain feedback for future improvements. Due to space constraints, we include 2 student responses. One student mentioned, "The simpleness, the animative way of explaining data structure is definitely the strength of DSAScratch". For weakness, only one student mentioned, "There could be some sort of pdf where all the blocks and function of the blocks are explained with examples." Since this was a short 3-day workshop we could accommodate only a few examples. In the future, we would prepare a detailed summary of these concepts with examples.

\subsection{\textbf{Summary of Evaluation:}} 
Pre-survey results indicate that students were novices and were interested in learning about programming section  \ref{presurvey}.
The results of DSAScratch's post-survey indicate that most students were satisfied with the experience. All 10 students unanimously agreed that the workshop gave them insight into data structures and algorithms. Among the 10 students, 9 (90\%) found DSAScratch easy to use, while 1 (10\%) found it difficult. In addition, 9 of the 10 students 
 (90\%) who attended the workshop stated they were likely to recommend DSAScratch to their peers, while 1 student remained neutral. Quiz-1 had 5 questions with each question carrying 1 point. On Quiz-1 8 (80\%) of the  participants got all the questions correct, whereas 2 (20\%) participants got 4 on 5. This indicates that the workshop helped most students grasp the fundamentals of Scratch.
Quiz-2 included 11 questions with 1 point each designed to test data structure and algorithms fundamentals. For Quiz-2 scores referring to Fig. \ref{fig:quiz2}, The results of Quiz-2 indicate that most students understood the basic concepts of data structures and algorithms covered during the workshop. 8 (80\%) students out of the 10 indicated that they were encouraged to use DSAScratch in the future while 2 (20\%) remained neutral. 

\subsection{\textbf{Evaluation of student solutions to Challenges}}

\begin{figure}[t]
  \centering
    \includegraphics[width=0.5\textwidth]{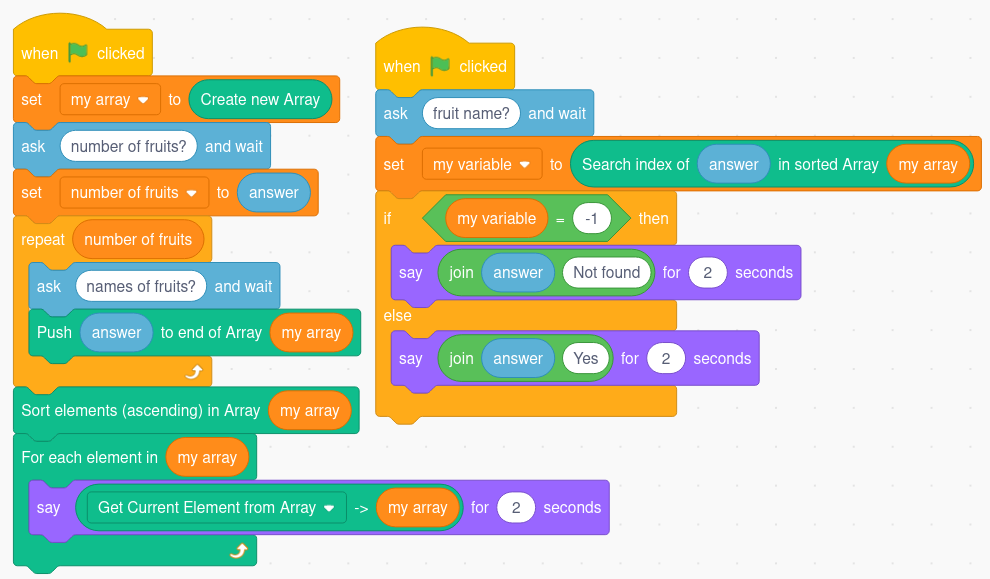}
    \caption{Correct student solution to Lab-1 challenge}
    \label{fig:correctsol}
\end{figure}

In this section, we analyze solutions submitted by students to the lab challenges to perform a quantitative analysis of DSAScratch. Out of the 3 challenges administered at the end of the workshop (refer Section \ref{quizzes} for challenges), 7 (70\%) students submitted correct solutions to all 3 challenges. Meanwhile, 3 (30\%) students submitted incorrect or partial solutions. For the first two challenges, we designed them to be easy, while for the third, we designed them to be of moderate difficulty. Most students had no prior knowledge of arrays, dictionaries, sets, searching, and sorting. In addition to helping students translate their learning into practice, these challenges were designed to evaluate the framework's simplicity. The first challenge in Section \ref{quizzes} requires students to use their knowledge of DSAScratch blocks but also demands familiarity with blocks from other basic constructs Scratch provides. The solution given by a student for the first challenge described in Section \ref{quizzes} consists of 2 parts, the challenge expects the students to accept inputs from the user and store them into a DSAScratch data structure and then perform a search for this item within the data structure. Additionally, the program must dynamically accept an arbitrary number of inputs from the user using Scratch input-output (I/O) blocks. A novice programmer may find this particularly challenging. It is important to understand how loops work to accomplish this. The student solution on the left side of Fig. \ref{fig:correctsol} shows that the student correctly applied loops to store a collection of items into a DSAScratch array data structure and combines this idea along with other Scratch programming constructs such as \texttt{ask} block to perform I/O. The student also demonstrates the usage of Scratch \texttt{repeat} block to store items. In the next step on the right side of Fig. \ref{fig:correctsol}, the student correctly uses a DSAScratch \texttt{for each element in} block to iterate over a collection of items inside the array variable named as \textit{my array}. Further, the student implements a linear search to find if the searched fruit exists inside \textit{my array} variable using a \textit{for each element in} block refer to Fig. \ref{fig:correctsol} right. Moreover, it confirms that the student understood Scratch basic constructs and their usage and combined it with the learning of DSAScratch blocks and concepts to provide solutions to lab challenges. 



\begin{figure}[t]
  \centering
    \includegraphics[width=0.5\textwidth]{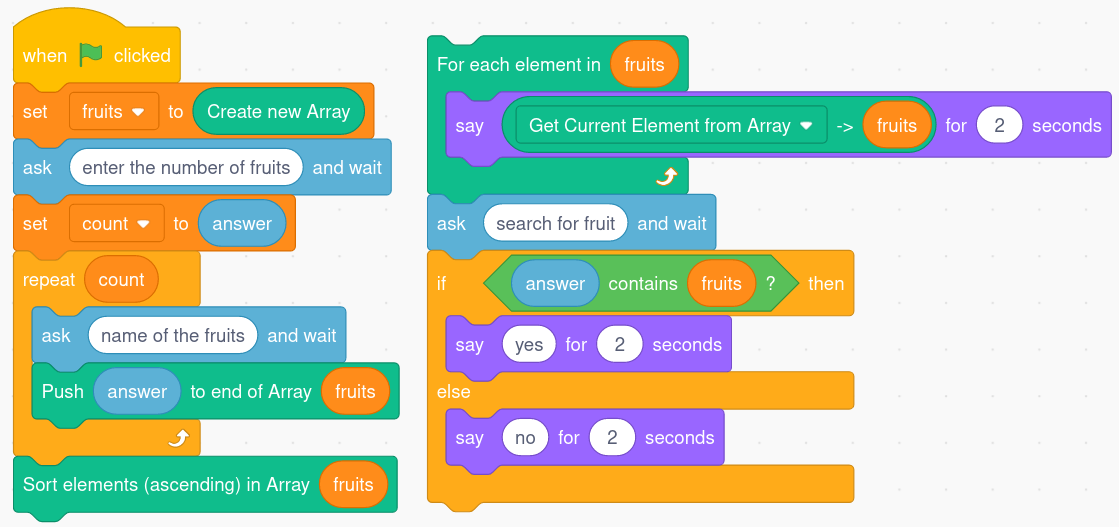}
    \caption{Incorrect student solution to Lab-1 challenge}
    \label{fig:incorrect}
\end{figure}

Next, Fig. \ref{fig:incorrect} shows a partially correct solution to Lab-1. Here the student followed the same pattern as in Fig. \ref{fig:correctsol}, storing items in an Array correctly using loops. However, the logic used to perform the search is incorrect. This may be due to the student being encouraged to use \textbf{contains ?} block from the operators' block category of Scratch to search inside the DSAScratch Array which is not possible. Due to the fact that Scratch \textbf{contains ?} is a block used to perform string matching given a text \textbf{contains ? } block matches words or letters. As an example, if the search text is "ball" and the search string is "football", then \textbf{contains ? } returns ‘’True’’. 

\section{Potential Benefits}

We show that the DSAScratch extension will be beneficial in providing an intermediate means of teaching advanced CS curricula using a block-based programming environment such as Scratch. According to research, block-based programming is taught in K-12 high schools to teach the basic concepts of computer science. Educators may find DSAScratch useful, while introducing advanced CS concepts when they divide subject matter into abstraction, logic, conditionals, loops, and writing basic algorithms\cite{csimportance1} as a part of introductory K-12 computer science curricula. Since visual interfaces like Scratch allow students to abstract, decompose problems, and build logic without focusing on technical details. Because Scratch is an open-source and free tool, using DSAScratch within Scratch can make these concepts accessible to a wide audience with minimal investment \cite{lowcost}. In introductory programming materials on Scratch, students will be able to learn these concepts more easily with the DSAScratch extension.



\section{Conclusion and Future work}

We designed the DSAScratch framework to simplify learning of data structures and algorithms, by abstracting common data structures and algorithms such as Arrays, Sets, Dictionaries, Searching, and Sorting. Furthermore, to evaluate DSAScratch's ability to teach intuitively, we designed a workshop and a user study to evaluate the  framework's ability to simplify complex computer science  concepts.
Through the workshop we introduced these concepts using the DSAScratch framework on Scratch, the results of the evaluations suggest that it  encouraged students to learn more about the topic post-workshop. We also emphasize that these initial results encourage further evaluation of DSAScratch's effectiveness. The framework will be evaluated in the future using more diverse student groups (e.g., sociodemographic factors, access to technology, and experience). It is also necessary to gather additional insights in order to improve  DSAScratch. Additionally, future developments will include data structures such as stacks, queues, trees, and graphs.



\bibliographystyle{IEEEtran}
\bibliography{main}

\begin{thebibliography}{10}
\providecommand{\url}[1]{#1}
\csname url@samestyle\endcsname
\providecommand{\newblock}{\relax}
\providecommand{\bibinfo}[2]{#2}
\providecommand{\BIBentrySTDinterwordspacing}{\spaceskip=0pt\relax}
\providecommand{\BIBentryALTinterwordstretchfactor}{4}
\providecommand{\BIBentryALTinterwordspacing}{\spaceskip=\fontdimen2\font plus
\BIBentryALTinterwordstretchfactor\fontdimen3\font minus
  \fontdimen4\font\relax}
\providecommand{\BIBforeignlanguage}[2]{{%
\expandafter\ifx\csname l@#1\endcsname\relax
\typeout{** WARNING: IEEEtran.bst: No hyphenation pattern has been}%
\typeout{** loaded for the language `#1'. Using the pattern for}%
\typeout{** the default language instead.}%
\else
\language=\csname l@#1\endcsname
\fi
#2}}
\providecommand{\BIBdecl}{\relax}
\BIBdecl

\bibitem{csimportance1}
S.~Grover and R.~Pea, \emph{Computational Thinking: A Competency Whose Time Has
  Come}, 2017.

\bibitem{csimportance2}
W.~Huang and C.-K. Looi, ``A critical review of literature on “unplugged”
  pedagogies in k-12 computer science and computational thinking education,''
  \emph{Computer Science Education}, 2021.

\bibitem{difficultyteaching2}
R.~P. Medeiros, G.~L. Ramalho, and T.~P. Falcão, ``A systematic literature
  review on teaching and learning introductory programming in higher
  education,'' \emph{IEEE Transactions on Education}, 2019.

\bibitem{difficultyteaching3}
\BIBentryALTinterwordspacing
S.~Su, E.~Zhang, P.~Denny, and N.~Giacaman, ``A game-based approach for
  teaching algorithms and data structures using visualizations,'' in
  \emph{Proceedings of the 52nd ACM Technical Symposium on Computer Science
  Education}, ser. SIGCSE '21.\hskip 1em plus 0.5em minus 0.4em\relax
  Association for Computing Machinery, 2021. [Online]. Available:
  \url{https://doi.org/10.1145/3408877.3432520}
\BIBentrySTDinterwordspacing

\bibitem{datastructureimportance1}
K.~L. Hart, ``Designing high school computer science curriculum to address
  real-world biological problems with computational thinking,'' 2021.

\bibitem{fredmartin}
\BIBentryALTinterwordspacing
D.~Touretzky, C.~Gardner-McCune, F.~Martin, and D.~Seehorn, ``Envisioning ai
  for k-12: What should every child know about ai?'' \emph{Proceedings of the
  AAAI Conference on Artificial Intelligence}, 2019. [Online]. Available:
  \url{https://ojs.aaai.org/index.php/AAAI/article/view/5053}
\BIBentrySTDinterwordspacing

\bibitem{difficultyteaching}
R.~Lawrence, ``Teaching data structures using competitive games,'' \emph{IEEE
  Transactions on Education}, 2004.

\bibitem{scratchhighschool2}
S.~Nikiforos, C.~Kontomaris, and K.~Chorianopoulos, ``Mit scratch: A powerful
  tool for improving teaching of programming,'' 2013.

\bibitem{cognition}
M.~Mladenović, z.~zanko, and M.~Aglić, ``The impact of using program
  visualization techniques on learning basic programming concepts at the k–12
  level,'' \emph{Computer Applications in Engineering Education}, 2021.

\bibitem{visuallanguagedatastructures}
C.-K. Chang, Y.-F. Yang, and Y.-T. Tsai, ``Exploring the engagement effects of
  visual programming language for data structure courses,'' \emph{Educ. Inf.},
  2017.

\bibitem{scratchpopular}
\BIBentryALTinterwordspacing
{Scratch Foundation}, ``Scratch - {Imagine}, {Program}, {Share}.'' [Online].
  Available: \url{https://scratch.mit.edu/}
\BIBentrySTDinterwordspacing

\bibitem{AIhigh}
J.~Estevez, G.~Garate, and M.~Graña, ``Gentle introduction to artificial
  intelligence for high-school students using scratch,'' \emph{IEEE Access},
  2019.

\bibitem{scratchwideadoption}
\BIBentryALTinterwordspacing
H.~Montiel and M.~G. Gomez-Zermeño, ``Educational challenges for computational
  thinking in k–12 education: A systematic literature review of “scratch”
  as an innovative programming tool,'' \emph{Computers}, 2021. [Online].
  Available: \url{https://www.mdpi.com/2073-431X/10/6/69}
\BIBentrySTDinterwordspacing

\bibitem{scratchimportant}
N.~Bean, J.~Weese, R.~Feldhausen, and R.~S. Bell, ``Starting from scratch:
  Developing a pre-service teacher training program in computational
  thinking,'' in \emph{2015 IEEE Frontiers in Education Conference (FIE)},
  2015.

\bibitem{activelearning}
\BIBentryALTinterwordspacing
M.~Wolff, M.~J. Wagner, S.~Poznanski, J.~Schiller, and S.~Santen, ``Not another
  boring lecture: Engaging learners with active learning techniques,''
  \emph{The Journal of Emergency Medicine}, 2015. [Online]. Available:
  \url{https://www.sciencedirect.com/science/article/pii/S0736467914009305}
\BIBentrySTDinterwordspacing

\bibitem{activelearning1}
\BIBentryALTinterwordspacing
S.~Popat and L.~Starkey, ``Learning to code or coding to learn? a systematic
  review,'' \emph{Computers \& Education}, 2019. [Online]. Available:
  \url{https://www.sciencedirect.com/science/article/pii/S0360131518302768}
\BIBentrySTDinterwordspacing

\bibitem{needfortoolsblktotext}
L.~Moors, A.~Luxton-Reilly, and P.~Denny, ``Transitioning from block-based to
  text-based programming languages,'' in \emph{2018 International Conference on
  Learning and Teaching in Computing and Engineering (LaTICE)}, 2018.

\bibitem{advancedconcepts2}
C.~G. von Wangenheim, J.~C.~R. Hauck, F.~S. Pacheco, and M.~F.~B. Bueno,
  ``Visual tools for teaching machine learning in k-12: A ten-year systematic
  mapping,'' \emph{Education and Information Technologies}, 2021.

\bibitem{cryptoscratch}
N.~Percival, P.~Rayavaram, S.~Narain, and C.~S. Lee, ``Cryptoscratch:
  Developing and evaluating a block-based programming tool for teaching k-12
  cryptography education using scratch,'' in \emph{2022 IEEE Global Engineering
  Education Conference (EDUCON)}, 2022.

\bibitem{scratchhighschool1}
\BIBentryALTinterwordspacing
I.~Ouahbi, F.~Kaddari, H.~Darhmaoui, A.~Elachqar, and S.~Lahmine, ``Learning
  basic programming concepts by creating games with scratch programming
  environment,'' \emph{Procedia - Social and Behavioral Sciences}, 2015.
  [Online]. Available:
  \url{https://www.sciencedirect.com/science/article/pii/S1877042815024842}
\BIBentrySTDinterwordspacing

\bibitem{bridgeprogram}
S.-M. Liao, ``Scratch to r: Toward an inclusive pedagogy in teaching coding,''
  \emph{Journal of Statistics and Data Science Education}, 2022.

\bibitem{learningtransferblocktotext}
\BIBentryALTinterwordspacing
D.~Bau, J.~Gray, C.~Kelleher, J.~Sheldon, and F.~Turbak, ``Learnable
  programming: Blocks and beyond,'' \emph{Communications of the ACM}, 2017.
  [Online]. Available: \url{https://doi.org/10.1145/3015455}
\BIBentrySTDinterwordspacing

\bibitem{scratchbeforepython}
M.~Dorling and D.~White, ``Scratch: A way to logo and python,''
  \emph{Proceedings of the 46th ACM Technical Symposium on Computer Science
  Education}, 2015.

\bibitem{novice1}
\BIBentryALTinterwordspacing
T.~W. Price and T.~Barnes, ``Comparing textual and block interfaces in a novice
  programming environment,'' in \emph{Proceedings of the Eleventh Annual
  International Conference on International Computing Education Research}, ser.
  ICER '15.\hskip 1em plus 0.5em minus 0.4em\relax Association for Computing
  Machinery, 2015. [Online]. Available:
  \url{https://doi.org/10.1145/2787622.2787712}
\BIBentrySTDinterwordspacing

\bibitem{scaffoldedlearning}
D.~Nam, Y.~Kim, and T.-S. Lee, ``The effects of scaffolding-based courseware
  for the scratch programming learning on student problem solving skill,'' in
  \emph{Proceedings of the 18th International Conference on Computers in
  Education}, 2010.

\bibitem{deepscratch}
W.~C. Gresse~von, R.~Hauck, Jean~C., F.~S. Pacheco, and F.~Bertonceli
  Bueno~Matheus, ``Visual tools for teaching machine learning in k-12: A
  ten-year systematic mapping,'' \emph{Education and Information Technologies},
  2021.

\bibitem{blocksforsensors}
J.~Judvaitis, A.~Elsts, and L.~Selavo, ``Demo abstract: Seal-blockly: Sensor
  network visual programming using a web browser,'' in \emph{10th European
  Conference of Wireless Sensor Networks}, 2013.

\bibitem{advancedconceptsscratch1}
J.~D.~R. García, J.~M. León, M.~R. González, and G.~Robles, ``Developing
  computational thinking at school with machine learning: An exploration,'' in
  \emph{2019 International Symposium on Computers in Education (SIIE)}, 2019.

\bibitem{lowcost}
\BIBentryALTinterwordspacing
K.~Yamamori, ``Classroom practices of low-cost stem education using scratch,''
  \emph{Journal of Advanced Research in Social Sciences and Humanities}, 2019.
  [Online]. Available: \url{http://dx.doi.org/10.2139/ssrn.3791166}
\BIBentrySTDinterwordspacing

\bibitem{difficultylearningDS}
\BIBentryALTinterwordspacing
D.~Zingaro, C.~Taylor, L.~Porter, M.~Clancy, C.~Lee, S.~Nam~Liao, and K.~C.
  Webb, ``Identifying student difficulties with basic data structures,'' in
  \emph{Proceedings of the 2018 ACM Conference on International Computing
  Education Research}, ser. ICER '18.\hskip 1em plus 0.5em minus 0.4em\relax
  New York, NY, USA: Association for Computing Machinery, 2018. [Online].
  Available: \url{https://doi.org/10.1145/3230977.3231005}
\BIBentrySTDinterwordspacing

\bibitem{dstblocks}
\BIBentryALTinterwordspacing
D.~F. Almanza-Cortés, M.~F. Del Toro-Salazar, R.~A. Urrego-Arias, P.~G.
  Feijóo-García, and F.~De~la Rosa-Rosero, ``Scaffolded block-based
  instructional tool for linear data structures: A constructivist design to
  ease data structures’ understanding,'' \emph{International Journal of
  Emerging Technologies in Learning (iJET)}, 2019. [Online]. Available:
  \url{https://online-journals.org/index.php/i-jet/article/view/10051}
\BIBentrySTDinterwordspacing

\bibitem{block4ds}
P.~G. Feijóo-García, S.~Wang, J.~Cai, N.~Polavarapu, C.~Gardner-McCune, and
  E.~D. Ragan, ``Design and evaluation of a scaffolded block-based learning
  environment for hierarchical data structures,'' in \emph{2019 IEEE Symposium
  on Visual Languages and Human-Centric Computing (VL/HCC)}, 2019.

\bibitem{Scratchmain}
\BIBentryALTinterwordspacing
M.~Resnick, J.~Maloney, A.~Monroy-Hern\'{a}ndez, N.~Rusk, E.~Eastmond,
  K.~Brennan, A.~Millner, E.~Rosenbaum, J.~Silver, B.~Silverman, and Y.~Kafai,
  ``Scratch: Programming for all,'' \emph{Communications of the ACM}, 2009.
  [Online]. Available: \url{https://doi.org/10.1145/1592761.1592779}
\BIBentrySTDinterwordspacing

\bibitem{atomiccomponentsprocedure}
\BIBentryALTinterwordspacing
S.~P. Rose, M.~P.~J. Habgood, and T.~Jay, ``Designing a programming game to
  improve children’s procedural abstraction skills in scratch,''
  \emph{Journal of Educational Computing Research}, 2020. [Online]. Available:
  \url{https://doi.org/10.1177/0735633120932871}
\BIBentrySTDinterwordspacing

\bibitem{blkcategories}
\BIBentryALTinterwordspacing
{Scratch Foundation}, ``Block categories, scratch wiki.'' [Online]. Available:
  \url{ttps://en.scratch-wiki.info/wiki/Block\_Categories}
\BIBentrySTDinterwordspacing

\bibitem{toee}
Y.~Park and Y.~Shin, ``Tooee: A novel scratch extension for k-12 big data and
  artificial intelligence education using text-based visual blocks,''
  \emph{IEEE Access}, 2021.

\bibitem{blockstypes}
\BIBentryALTinterwordspacing
{Scratch Foundation}, ``Blocks, scratch wiki.'' [Online]. Available:
  \url{https://en.scratch-wiki.info/wiki/Blocks}
\BIBentrySTDinterwordspacing

\bibitem{es6}
\BIBentryALTinterwordspacing
{W3 Schools}, ``Ecmascript 6.'' [Online]. Available:
  \url{https://www.w3schools.com/Js/js\_es6.asp}
\BIBentrySTDinterwordspacing

\end{thebibliography}

\end{document}